\documentstyle[12pt]{article}
\begin{document}
\newcommand{\eq}{\begin{equation}}
\newcommand{\en}{\end  {equation}}

\begin{titlepage}

\title{
The Spin Structure of the $qq$ Interaction and
the Mass Spectra of Bound $q\bar{q}$ Systems: \\
Different Versions of the 3-Dimensional
Reductions of the Bethe-Salpeter Equation
}

\author{
T.Babutsidze$^1$, T.Kopaleishvili$^1$ and A.Rusetsky$^{1,2}$\\ \\
$^1${\em High Energy Physics Institute, Tbilisi State University,}\\
{\em University St 9, 380086, Tbilisi, Republic of Georgia}\\
$^2${\em Bogoliubov Laboratory of Theoretical Physics,}\\
{\em Joint Institute for Nuclear Research,}\\
{\em 141980 Dubna (Moscow region), Russia}\\
}

\maketitle

\abstract
\baselineskip 20pt
Bound $q\bar{q}$-systems are considered in the framework of three
different versions of the 3-dimensional reduction of the Bethe-Salpeter
equation, all having the correct one-body limit when one of the constituent
quark masses tends to infinity, and in the framework of the Salpeter
equation. The spin structure of confining $qq$ interaction potential
is taken in the form $x\gamma_{1}^{0}\gamma_{2}^{0}+(1-x)I_{1}I_{2}$,
with $0\leq x \leq 1$. The problem of existence (nonexistence) of stable
solutions of 3-dimensional relativistic equations for bound $q\bar{q}$
systems is studied for different values of $x$ from this interval.
Some other aspects of this problem are discussed.

\begin{center}
\end{center}

\end{titlepage}

\newpage

As it is well-known, at present the spin (Lorentz) structure of the $qq$
interaction is not established theoretically in QCD,
a fundamental theory of strong interactions.
Consequently, it is interesting to consider different possible choices
for the spin structure, as it was done in Refs. \cite{FBS}-\cite{Olson}
where the bound $q\bar{q}$ systems in the framework of the Salpeter equation were
investigated. Further, it is well known that the Salpeter equation is
the simplest version of the 3-dimensional (3D) reduction of the
Bethe-Salpeter (BS) equation, the latter being believed to provide a natural
basis to study bound $q\bar{q}$ systems in the framework of
the Constituent Quark Model. Namely, the Salpeter equation is
obtained from the BS equation in a straightforward way, when the kernel of the latter is
assumed to have the instantaneous (static) form. However, in the instantaneous
approximation there exist other possible versions \cite{MW,CJ} of the
3D reduction of the BS equation which, unlike the Salpeter equation,
have a correct one-body limit when the mass of one of the constituents
tends to infinity. These versions, which can be derived by choosing
appropriate effective 3D-Green's function for two noninteracting fermions,
will hereafter be referred to as the MW and CJ versions, respectively.
Moreover, a new version of the effective propagator for two free scalar particles,
guaranteeing the existence of the correct one-body limit for 3D-equations, was
suggested in Ref. \cite{MNK}. The effective 3D-Green's function for two
noninteracting fermions can be constructed along the lines similar to
Ref. \cite{MNK}, in a standard manner (see below).

Taking into account the fact that the relativistic effects are
important for $q\bar{q}$ systems containing two light quarks,
as well as for heavy-light systems, it seems interesting to carry out
the investigation of this sort of systems in the framework of the above mentioned
3D relativistic equations and study the dependence of the properties of
the bound $q\bar{q}$ systems on the spin (Lorentz) structure of
the confining part of $qq$-interaction. In the present report, we
deal with this problem as concerned to the $q\bar q$ mass spectrum.

The relativistic 3D equations for the wave function of bound
$q\bar{q}$ systems, corresponding to the instantaneous (static)
BS kernel ($K(P;p,p')\rightarrow K_{st}(\vec{p},\vec{p}^{~'})$),
for all versions considered below can be written in a common form
(in the cm frame)
\eq
        \tilde{\Phi}_ {M}(\vec{p}) =
        \tilde{G}_{0eff}(M,\vec{p})
        \int\frac{d\vec{p}^{~'}}{(2\pi)^{3}}
         [iK_{st}(\vec{p},\vec{p}^{~'}) \equiv
        \hat{V}(\vec{p},\vec{p}^{~'})]
        \tilde{\Phi}_{M}(\vec{p}^{~'})
\en
\noindent
where $M$ is the bound system mass. The equal-time wave function
$\tilde{\Phi}_{M}(\vec{p})$ is related to the BS amplitude $\Phi_{P}(p)$ by
\eq
       \tilde{\Phi}_{M}(\vec{p}) =
       \int\frac{dp_{0}}{2\pi}\,\Phi_{P=(M,\vec{0})}(p)
\en
\noindent
and the effective 3D Green's function of the two noninteracting quark system
$\tilde{G}_{0eff}$ is defined as
\eq
\tilde{G}_{0eff}(M,\vec{p})=\int\frac{dp_0}{2\pi i}\,
[G_{0eff}(M,p) = g_{0eff}(M,p)(\not\! p_{1}+m_{1})(\not\! p_{2}+m_{2})]
\en
\noindent
Here, $G_{0eff}$ is the effective two fermion free propagator and
$g_{0eff}$ is the effective propagator of two scalar particles. The operator
$\tilde{G}_{0eff}$ can be given in the form
\eq
\tilde{G}_{0eff}(M,\vec{p}) =
\sum\limits_{\alpha_{1}=\pm}\sum\limits_{\alpha_{2}=\pm}\,
\frac{D^{(\alpha_{1}\alpha_{2})}(M,p)}{d(M,p)}\,\,
\Lambda_{12}^ {(\alpha_1\alpha_2)}
(\vec{p},-\vec{p})\,\,\gamma_{1}^{0}\gamma_{2}^{0},\,\,\,\,\,\,\,\,
p\equiv|\vec{p}|
\en
\noindent
where the projection operators $\Lambda_{12}^{(\alpha_1\alpha_2)}$
are defined as
$$
 \Lambda_{12}^{(\alpha_1\alpha_2)}(\vec{p}_{1},\vec{p}_{2})
=\Lambda_{1}^{(\alpha_1)}(\vec{p}_{1})\otimes
 \Lambda_{2}^{(\alpha_2)}(\vec{p}_{2}),\,\,\,\,\,
 \Lambda_{i}^{(\alpha_i)}(\vec{p}_{i})
=\frac{(\omega_{i}+\hat h_{i}(\vec{p}_{i}))}{2\omega_{i}}
$$
\eq
\hat h_{i}(\vec{p}_{i})=\vec{\alpha}_{i}\vec{p}_{i}+m_{i}\gamma_{i}^{0},
\,\,\,\,\,\,\,\,
\omega_{i}=(m_{i}^{2}+\vec{p}_{i}^{~2})^{1/2}
\en
\noindent
and the functions $D^{(\alpha_{1}\alpha_{2})}(M,p)$ and $d(M,p)$ are given by
\begin{eqnarray}
D^{(\alpha_{1}\alpha_{2})}(M,p)=
\frac{(-1)^{\alpha_{1}+\alpha_{2}}}
{\omega_{1}+\omega_{2}-(\alpha_{1}E_{1}+\alpha_{2}E_{2})},
\,\,\,\,\,
d(M,p)=1
&\nonumber\\[2mm]
E_{1}+E_{2}=M,\,\,\,\,
E_{1}-E_{2}=\frac{m_{1}^{2}-m_{2}^{2}}{M}\equiv b_0
&{\mbox{(MW version)}}\\[2mm]
D^{(\alpha_{1}\alpha_{2})}(M,p)=
(E_{1}+\alpha_{1}\omega_{1})(E_{2}+\alpha_{2}\omega_{2})
&\nonumber\\[2mm]
d(M,p)=2(\omega_{1}+\omega_{2}),\,\,\,\,a=
E_{i}^{2}-\omega_{i}^{2}=[M^{2}+b_{0}^{2}-2(\omega_{1}^{2}+\omega_{2}^{2})]/4
&\mbox{(CJ version)}
\end{eqnarray}
\noindent
(see Ref. \cite{TT} where the mass spectrum of the bound $q\bar{q}$ systems
was investigated in the framework of the MW and CJ versions of the relativistic
3D equations in the configuration space).

Note that expression (7) was derived from Eq. (3) by using that
for $g_{0eff}(M,p)$ determined from the dispersion relation.
The same relation is satisfied by the expression of $g_{0eff}(M,p)$
suggested in Ref. \cite{MNK} (see formula (10) therein).
According to the prescription given in Ref. \cite{MNK}, particles in
the intermediate states are allowed to go off shell proportionally to
their mass, so that when one of the particles becomes infinitely massive,
it automatically is kept on mass shell and the corresponding equation
can be reduced to the one-body equation. Using this expression for
$g_{0eff}(M,p)$ from Eq. (3) we obtain the expression for $\tilde{G}_{0eff}$
given by Eq. (4). (Note that formula (11) in Ref. \cite{MNK} is not
correct as it does not follow from formula (10)). So we obtain
$$
D^{(\alpha_{1}\alpha_{2})}(M,p)=
(E_{1}+\alpha_{1}\omega_{1})(E_{2}+\alpha_{2}\omega_{2})-\frac{R-b}{2y}
\biggl[\frac{R-b}{2y}
+(E_{1}+\alpha_{1}\omega_{1})-(E_{2}+\alpha_{2}\omega_{2})\biggr]
$$
$$
d(M,p)=2RB,
\hspace*{0.5cm}
R=(b^{2}-4y^{2}a)^{1/2},
\hspace*{0.5cm}
B=\frac{R-b}{2y}\biggl[\frac{R-b}{2y}+b_{0}\biggr]+a,
$$
\eq
b=M+b_{0}y,
\hspace*{0.5cm}
y=\frac{m_{1}-m_{2}}{m_{1}+m_{2}}
\en
This version will below be referred to as the MNK version.

Using the properties of the projection operators
$\Lambda_{12}^ {(\alpha_1\alpha_2)}$ and formulae (4-8),
the following system of equations can be derived from Eq. (1)
$$
[M-(\alpha_{1}\omega_{1}+\alpha_{2}\omega_{2})]
\tilde{\Phi}_ {M}^{(\alpha_{1}\alpha_{2})}(\vec{p})=
$$
\eq
=A^{(\alpha_{1}\alpha_{2})}(M,p)\,
\Lambda_{12}^{(\alpha_1\alpha_2)}(\vec{p},-\vec{p})
\int\frac{d\vec{p}^{~'}}{(2\pi)^{3}}\,
\gamma_{1}^{0}\gamma_{2}^{0}\,
\hat{V}(\vec{p},\vec{p}^{~'})
\sum\limits_{\alpha_{1}^{'}=\pm}\,\sum\limits_{\alpha_{2}^{'}=\pm}\,
\tilde{\Phi}_ {M}^{(\alpha_{1}^{'}\alpha_{2}^{'})}(\vec{p}^{~'})
\en
\noindent
where $\tilde{\Phi}_ {M}^{(\alpha_{1}\alpha_{2})}(\vec{p})=
\Lambda_{12}^{(\alpha_1\alpha_2)}(\vec{p},-\vec{p})\,
\tilde{\Phi}_ {M}(\vec{p})$
and the functions $A^{(\alpha_{1}\alpha_{2})}$ are given by
\begin{eqnarray}
A^{(\pm\pm)}=\pm 1,
\hspace*{1.cm}
A^{(\pm\mp)}=\frac{M}{\omega_{1}+\omega_{2}}
\hspace*{1.2cm}
\mbox{(MW version)}&
\\[2mm]
A^{(\alpha_{1}\alpha_{2})}=
\frac{M+(\alpha_{1}\omega_{1}+\alpha_{2}\omega_{2})}
{2(\omega_{1}+\omega_{2})}
\hspace*{1.2cm}
\mbox{(CJ version)}&
\\[2mm]
A^{(\alpha_{1}\alpha_{2})}=\frac{1}{2RB}\biggl\{
a[M+(\alpha_{1}\omega_{1}+\alpha_{2}\omega_{2})] - \hspace*{4.2cm} &
\nonumber\\[2mm]
-[M-(\alpha_{1}\omega_{1}+\alpha_{2}\omega_{2})]
\frac{R-b}{2y}
\biggl[\frac{R-b}{2y}+(E_{1}+\alpha_{1}\omega_{1})-
(E_{2}+\alpha_{2}\omega_{2})\biggr]\biggr\}
\hspace*{1.2cm}
\mbox{(MNK version)}&
\end{eqnarray}
\par
As to the Salpeter equation, it can be obtained from the MW version
by putting $A^{(\pm\mp)}=0$ and $\tilde{\Phi}_{M}^{(\pm\mp)}=0$.

Further, we write the unknown function
$\tilde{\Phi}_{M}^{(\alpha_{1}^{'}\alpha_{2}^{'})}$
in Eq. (9) in the form analogous to that used in Ref. \cite{4A},
where the bound $q\bar{q}$ systems were studied in the framework of
the Salpeter equation
\eq
\tilde{\Phi}_ {M}^{(\alpha_{1}\alpha_{2})}(\vec{p})=
N_ {12}^{(\alpha_{1}\alpha_{2})}(p)
\pmatrix{1\cr
 \alpha_{1}\vec{\sigma}_{1}\vec{p}/(\omega_{1}+\alpha_{1}m_{1})
}
\otimes
\pmatrix{1\cr
-\alpha_{2}\vec{\sigma}_{2}\vec{p}/(\omega_{2}+\alpha_{2}m_{2})
}
\hspace*{0.3cm}
\chi_{M}^{(\alpha_{1}\alpha_{2})} (\vec{p})
\en
\noindent where
\eq
N_ {12}^{(\alpha_{1}\alpha_{2})}(p)=
\biggl(\frac{\omega_{1}+\alpha_{1}m_{1}}{2\omega_{1}}\biggr)^{1/2}
\biggl(\frac{\omega_{2}+\alpha_{2}m_{2}}{2\omega_{2}}\biggr)^{1/2}
\equiv N_ {1}^{(\alpha_{1})}(p)N_ {2}^{(\alpha_{2})}(p)
\en
Then, if the $qq$ interaction potential operator
$\hat{V}(\vec{p},\vec{p}^{~'})$ is taken in the form \cite{4A}
\eq
\hat{V}(\vec{p},\vec{p}^{~'})=\gamma_{1}^{0}\gamma_{2}^{0}
\hat{V}_{og}(\vec{p} - \vec{p}^{~'})+
[x\gamma_{1}^{0}\gamma_{2}^{0}+(1-x)I_{1}I_{2}]
\hat{V}_{c}(\vec{p} - \vec{p}^{~'}),
\hspace*{0.8cm}
(0\leq x \leq1),
\en
the following system of equations for the Pauli $2\otimes2$ wave
functions $\chi_{M}^{(\alpha_{1}\alpha_{2})}$  can be derived
$$
[M-(\alpha_{1}\omega_{1} + \alpha_{2}\omega_{2})]
\chi_{M}^{(\alpha_{1}\alpha_{2})}(\vec{p})=
$$
\eq
=A^{(\alpha_{1}\alpha_{2})}(M;p)
\sum\limits_{\alpha_{1}^{'}=\pm}\,\sum\limits_{\alpha_{2}^{'}=\pm}\,
\int\frac{d\vec{p}^{~'}}{(2\pi)^{3}}\,
\hat{V}_{eff}^{(\alpha_{1}\alpha_{2},\alpha_{1}^{'}\alpha_{2}^{'})}
(\vec{p},\vec{p}^{~'},\vec{\sigma}_{1},\vec{\sigma}_{2})
\chi_{M}^{(\alpha_{1}^{'}\alpha_{2}^{'})}(\vec{p}^{~'})
\en
\noindent
where the effective  $qq$ interaction operator
$\hat{V}_{eff}^{(\alpha_{1}\alpha_{2},\alpha_{1'}\alpha_{2'})}
(\vec{p},\vec{p}^{'},\vec{\sigma}_{1},\vec{\sigma}_{2})$
is expressed via the po\-ten\-ti\-als $V_{og}$ and $V_{c}$ and
some functions, taking account of relativistic kinematics. On using the
partial-wave expansion
\eq
\chi_{M}^{(\alpha_{1}\alpha_{2})}(\vec{p})=
\sum\limits_{LSJM_{J}}<\vec{n}|{LSJM_{J}}>\,\,
R_{LSJ}^{(\alpha_{1}\alpha_{2})}(\vec{p}),
\hspace*{1.8cm}
(\vec{n}=\vec{p}/p)
\en
From Eq. (16) the system of integral equations for the radial wave functions
$R_{LSJ}^{(\alpha_{1}\alpha_{2})}(p)$ can be obtained.

At the first stage our investigation is aimed at
comparative analysis of the different versions of the 3D relativistic
equations as concerns the existence of stable solutions
for different values of the vector-scalar mixing parameter $x$ in Eq. (15).
For this reason, in Eq. (15) we neglect the one-gluon exchange potential
and take the confining potential $V_{c}(r)$ in the oscillator form used in
Ref. \cite{2A}, which is a simplified though justified version
(for the light and light-heavy sectors) of a more general form used in
Ref. \cite{4A}. Namely, we take
$$
V_{c}(r)=\frac{4}{3}\alpha_{s}(m_{12}^{2})
\biggl(\frac{\mu_{12}\omega_{0}^{2}}{2}r^{2}-V_{0}\biggr)
$$
\eq
\mu_{12}=\frac{m_{1}m_{2}}{m_{12}},
\hspace*{0.5cm}
m_{12}=m_{1}+m_{2},
\hspace*{0.5cm}
\alpha_{s}(Q^{2})=\frac{12\pi}{33-2n_{f}}
\biggl(\mbox{ln}\frac{Q^{2}}{\Lambda^{2}}\biggr)^{-1}
\en
In the momentum space, the system of integral equations for the
radial functions $R_{LSJ}^{(\alpha_{1}\alpha_{2})}(p)$ with the above
potential is reduced to the system of second-order differential equations.
The solution of this equation is written in the form similar to that
given in Refs. \cite{4A,2A}
\eq
R_{LSJ}^{(\alpha_{1}\alpha_{2})}(p)=
\sum\limits_{n=0}^{\infty}C_{LSJn}^{(\alpha_{1}\alpha_{2})}
R_{nL}^{(\alpha_{1}\alpha_{2})}(p)
\en
\noindent
where $R_{nL}(p)$ are the well-known oscillator wave functions. Then, the
system of equations for $R_{LSJ}^{(\alpha_{1}\alpha_{2})}(p)$ is reduced
to the system of linear algebraic equations for the coefficients
$C_{LSJ}^{(\alpha_{1}\alpha_{2})}$.
\eq
M\,\, C_{LSJn}^{(\alpha_{1}\alpha_{2})}=
\sum\limits_{\alpha_{1}^{'},\alpha_{2}^{'}}
\sum\limits_{L^{'}S^{'}n^{'}}
H^{(\alpha_{1}\alpha_{2},\alpha_{1}^{'}\alpha_{2}^{'})}_{LSJn,L^{'}S^{'}L^{'}n^{'}}(M)
\,\,
C_{L^{'}S^{'}J^{'}n^{'}}^{(\alpha_{1}^{'}\alpha_{2}^{'})}
\en

Here it is necessary to stress that the matrix $H$
explicitly depends (except the Salpeter version) on the meson mass $M$ we
are looking for. Concequently, the system of equations (20) is nonlinear in $M$.

By truncating the sum in (19) at some fixed value $N_{max}$, the
eigenvalues $M$ and the corresponding coefficients
$C_{LSJn}^{(\alpha_{1}\alpha_{2})}$ can be determined from the system
of algebraic equations with the dimension $4(N_{max}+1)$
(for the Salpeter case we have $2(N_{max} +1)$ equations),
provided the procedure converges with the increase of
$N_{max}$. If the procedure does not converge, we interpret this as absence
of stable solutions to the initial equations. As it has been mentioned above, the
system  of equations (20) is nonlinear in $M$ for the MW, CJ and MNK
versions of the 3D equations. For the solution we use the iteration procedure:
at the first step, in the r.h.s. of Eq. (20) determining $M$, we substitute
the solution of the Salpeter equation and then find $M$ by the procedure
explained above. At the second step, we substitute the obtained solution
into the r.h.s. of Eq. (20) and the iterations are continued until the
result converges.

In the present paper, we calculate
the masses for the following $q\bar{q}$ systems with nonequal mass quarks:
$d\bar{s}$ ($^{1}S_{0}$, $^{3}S_{1}$, $^{1}P_{1}$, $^{3}P_{0}$, $^{3}P_{1}$,
$^{3}P_{2}$, $^{1}D_{2}$, $^{3}D_{1}$, $^{3}D_{3}$),
$c\bar{u}$ and $c\bar{s}$
($^{1}S_{0}$, $^{3}S_{1}$, $^{1}P_{1}$, $^{3}P_{2}$), for which the
values of the meson masses are known.
In the calculations the following values of the parameters of
the $qq$ -interaction potential (18) were used \cite{4A,2A}:
$\omega_{0} = 710~MeV$, $V_{0} = 525~MeV$, $\Lambda=120~MeV$,
$m_{u}=m_{d}=280~MeV$, $m_{s}=400~MeV$, $m_{c}=1470~MeV$.

On the basis of calculation of mass spectra of the above
$q\bar{q}$ systems we have arrived at the following conclusions:

The stable solutions of the MW, CJ and MNK versions of the 3D relativistic
equations always exist for $x=0$. For $x=1$ these solutions do not exist
for the majority of states under consideration. For the Salpeter
equation the situation is just opposite: for $x=1$ stable solutions
always exist whereas for $x=0$ the solutions do not exist for the majority
of the states studied. This agrees with the results obtained earlier
in Refs. \cite{FBS}-\cite{Olson} for the
$q\bar{q}$ systems of equal mass quarks from the light quark sector (u,d,s).
Moreover, for the CJ and MNK versions stable solutions always exist for
$(0\leq x \leq0.5)$. As to the MW version, this sort of solutions exist only for
$q\bar{q}$ systems with one heavy quark, whereas in order to provide
the existence of stable solutions in the same interval of $x$ for
light $q\bar{q}$ systems it is necessary to accept a much smaller
value for the confining potential strength parameter $\omega_{0}$
(e.g. $450~MeV$ instead of $710~MeV$). As to the interval
$(0.5\leq$x$\leq1)$, the existence (nonexistence) of stable solutions in
the MW, CJ and MNK versions depends on the quark sector (light or heavy),
quantum numbers of the $q\bar{q}$ system and on the value of the
parameter $\omega_{0}$. For the case of the Salpeter equation, the
situation again is  opposite - stable solutions always exist in the
interval $(0.5\leq x \leq1)$ for $q\bar{q}$ systems with both quarks from
light-quark sector \cite{FBSA}-\cite{Olson}, in the whole interval
$(0\leq$x$\leq1)$ for $q\bar{q}$ systems with both quarks from heavy-quark
sector ($c\bar{c}$) \cite{4A}, or from heavy-light sector (present result).
The existence of stable solutions of the relativistic equations under
consideration is mainly related to the presence of the
mixed ($+-$ and $-+$) energy components of the wave functions
in the equation for the $q\bar{q}$ bound state.

To illustrate these conclusions, in Tables 1. a,b,c  we give
the results of numerical solution of the system of equations (20) for
$d\bar{s}$, $c\bar{u}$ and $c\bar{s}$ bound systems for the states
$^{1}S_{0}$, $^{3}S_{1}$ and $^{1}P_{1}$.

Note that in order to obtain stable solutions of the 3D relativistic equations,
it is sufficient to take $N_{max}=4 - 7$ in the series (19).
This property is common for all 3D versions and meson states under
consideration.

A more detailed analysis (and the comparison with experiment) of the results
in the fra\-me\-work of the above considered versions of 3D equations with
the confining potential (18) (including the regularization problem of the
wave function normalization condition in the CJ and MNK versions), as well as
the description of decay properties of pseudoscalar and vector mesons
($P\rightarrow\mu\bar{\nu}$, $V\rightarrow e^{+}e^{-}$), will be published
separately.
\vspace*{.2cm}

{\it Acknowledgments.} One of the authors (A.R.) acknowledges the financial
support from Russian Foundation for Basic Research under contract
96-02-17435-a.

\newpage
Table 1.a  The mass spectrum (in GeV) for $(d\bar{s})$ system
\vspace*{0.7cm}

\begin{tabular}{|c|r|c|c|c|c|c|c|c|c|}
\hline
States&Versions&$\alpha_{1}\alpha_{2}$& x=0.0 & x=0.1 & x=0.3 & x=0.5 & x=0.7
& x=0.9 & x=1.0  \\
\hline
& \multicolumn{1}{c|}{MW$^{+}$}&\multicolumn{1}{c|}{++}&
\multicolumn{1}{c|}{0.826}&
\multicolumn{1}{c|}{0.836}&
\multicolumn{1}{c|}{0.854}&
\multicolumn{1}{c|}{0.870}&
\multicolumn{1}{c|}{0.884}&
\multicolumn{1}{c|}{0.895}&
\multicolumn{1}{c|}{0.900}\\
&\multicolumn{1}{c|}{"}&\multicolumn{1}{c|}{all}&
\multicolumn{1}{c|}{0.849}&
\multicolumn{1}{c|}{0.855}&
\multicolumn{1}{c|}{0.888}&
\multicolumn{1}{c|}{0.877}&
\multicolumn{1}{c|}{0.889}&
\multicolumn{1}{c|}{*}&
\multicolumn{1}{c|}{*}\\
\cline{2-10}
& \multicolumn{1}{c|}{CJ}&\multicolumn{1}{c|}{++}&
\multicolumn{1}{c|}{0.966}&
\multicolumn{1}{c|}{0.997}&
\multicolumn{1}{c|}{1.055}&
\multicolumn{1}{c|}{1.107}&
\multicolumn{1}{c|}{1.155}&
\multicolumn{1}{c|}{1.197}&
\multicolumn{1}{c|}{1.216}\\
&\multicolumn{1}{c|}{"}&\multicolumn{1}{c|}{all}&
\multicolumn{1}{c|}{1.025}&
\multicolumn{1}{c|}{1.042}&
\multicolumn{1}{c|}{1.078}&
\multicolumn{1}{c|}{1.117}&
\multicolumn{1}{c|}{1.160}&
\multicolumn{1}{c|}{*}&
\multicolumn{1}{c|}{*}\\
\cline{2-10}
 $^{1}S_{0}$&\multicolumn{1}{c|}{MNK}&
\multicolumn{1}{c|}{++}&
\multicolumn{1}{c|}{0.938}&
\multicolumn{1}{c|}{0.966}&
\multicolumn{1}{c|}{1.016}&
\multicolumn{1}{c|}{1.060}&
\multicolumn{1}{c|}{1.098}&
\multicolumn{1}{c|}{1.131}&
\multicolumn{1}{c|}{1.146}\\
&\multicolumn{1}{c|}{"}&\multicolumn{1}{c|}{all}&
\multicolumn{1}{c|}{0.990}&
\multicolumn{1}{c|}{1.005}&
\multicolumn{1}{c|}{1.036}&
\multicolumn{1}{c|}{1.069}&
\multicolumn{1}{c|}{*}&
\multicolumn{1}{c|}{*}&
\multicolumn{1}{c|}{*}\\
\cline{2-10}
& \multicolumn{1}{c|}{Sal}&\multicolumn{1}{c|}{++}&
\multicolumn{1}{c|}{0.951}&
\multicolumn{1}{c|}{0.981}&
\multicolumn{1}{c|}{1.035}&
\multicolumn{1}{c|}{1.082}&
\multicolumn{1}{c|}{1.124}&
\multicolumn{1}{c|}{1.161}&
\multicolumn{1}{c|}{1.178}\\
&\multicolumn{1}{c|}{"}&\multicolumn{1}{c|}{all}&
\multicolumn{1}{c|}{*}&
\multicolumn{1}{c|}{*}&
\multicolumn{1}{c|}{1.026}&
\multicolumn{1}{c|}{1.079}&
\multicolumn{1}{c|}{1.121}&
\multicolumn{1}{c|}{1.160}&
\multicolumn{1}{c|}{1.172}\\
\hline
& \multicolumn{1}{c|}{MW$^{+}$}&\multicolumn{1}{c|}{++}&
\multicolumn{1}{c|}{0.833}&
\multicolumn{1}{c|}{0.843}&
\multicolumn{1}{c|}{0.861}&
\multicolumn{1}{c|}{0.877}&
\multicolumn{1}{c|}{0.891}&
\multicolumn{1}{c|}{0.902}&
\multicolumn{1}{c|}{0.907}\\
&\multicolumn{1}{c|}{"}&\multicolumn{1}{c|}{all}&
\multicolumn{1}{c|}{0.855}&
\multicolumn{1}{c|}{0.861}&
\multicolumn{1}{c|}{0.873}&
\multicolumn{1}{c|}{0.877}&
\multicolumn{1}{c|}{0.903}&
\multicolumn{1}{c|}{*}&
\multicolumn{1}{c|}{*}\\
\cline{2-10}
& \multicolumn{1}{c|}{CJ}&\multicolumn{1}{c|}{++}&
\multicolumn{1}{c|}{0.985}&
\multicolumn{1}{c|}{1.017}&
\multicolumn{1}{c|}{1.076}&
\multicolumn{1}{c|}{1.129}&
\multicolumn{1}{c|}{1.177}&
\multicolumn{1}{c|}{1.220}&
\multicolumn{1}{c|}{1.239}\\
&\multicolumn{1}{c|}{"}&\multicolumn{1}{c|}{all}&
\multicolumn{1}{c|}{1.036}&
\multicolumn{1}{c|}{1.055}&
\multicolumn{1}{c|}{1.095}&
\multicolumn{1}{c|}{1.140}&
\multicolumn{1}{c|}{1.192}&
\multicolumn{1}{c|}{*}&
\multicolumn{1}{c|}{*}\\
\cline{2-10}
 $^{3}S_{1}$&\multicolumn{1}{c|}{MNK}&
\multicolumn{1}{c|}{++}&
\multicolumn{1}{c|}{0.958}&
\multicolumn{1}{c|}{0.986}&
\multicolumn{1}{c|}{1.035}&
\multicolumn{1}{c|}{1.079}&
\multicolumn{1}{c|}{1.117}&
\multicolumn{1}{c|}{1.150}&
\multicolumn{1}{c|}{1.165}\\
&\multicolumn{1}{c|}{"}&\multicolumn{1}{c|}{all}&
\multicolumn{1}{c|}{1.004}&
\multicolumn{1}{c|}{1.020}&
\multicolumn{1}{c|}{1.054}&
\multicolumn{1}{c|}{1.090}&
\multicolumn{1}{c|}{1.130}&
\multicolumn{1}{c|}{*}&
\multicolumn{1}{c|}{*}\\
\cline{2-10}
& \multicolumn{1}{c|}{Sal}&\multicolumn{1}{c|}{++}&
\multicolumn{1}{c|}{0.971}&
\multicolumn{1}{c|}{1.001}&
\multicolumn{1}{c|}{1.054}&
\multicolumn{1}{c|}{1.102}&
\multicolumn{1}{c|}{1.143}&
\multicolumn{1}{c|}{1.181}&
\multicolumn{1}{c|}{1.198}\\
&\multicolumn{1}{c|}{"}&\multicolumn{1}{c|}{all}&
\multicolumn{1}{c|}{*}&
\multicolumn{1}{c|}{*}&
\multicolumn{1}{c|}{1.037}&
\multicolumn{1}{c|}{1.095}&
\multicolumn{1}{c|}{1.141}&
\multicolumn{1}{c|}{1.181}&
\multicolumn{1}{c|}{1.198}\\
\hline
& \multicolumn{1}{c|}{MW$^{+}$}&\multicolumn{1}{c|}{++}&
\multicolumn{1}{c|}{1.045}&
\multicolumn{1}{c|}{1.059}&
\multicolumn{1}{c|}{1.085}&
\multicolumn{1}{c|}{1.106}&
\multicolumn{1}{c|}{1.124}&
\multicolumn{1}{c|}{1.139}&
\multicolumn{1}{c|}{1.145}\\
&\multicolumn{1}{c|}{"}&\multicolumn{1}{c|}{all}&
\multicolumn{1}{c|}{1.082}&
\multicolumn{1}{c|}{1.090}&
\multicolumn{1}{c|}{1.104}&
\multicolumn{1}{c|}{1.118}&
\multicolumn{1}{c|}{*}&
\multicolumn{1}{c|}{1.140}&
\multicolumn{1}{c|}{*}\\
\cline{2-10}
& \multicolumn{1}{c|}{CJ}&\multicolumn{1}{c|}{++}&
\multicolumn{1}{c|}{1.258}&
\multicolumn{1}{c|}{1.302}&
\multicolumn{1}{c|}{1.380}&
\multicolumn{1}{c|}{1.448}&
\multicolumn{1}{c|}{1.505}&
\multicolumn{1}{c|}{1.155}&
\multicolumn{1}{c|}{1.577}\\
&\multicolumn{1}{c|}{"}&\multicolumn{1}{c|}{all}&
\multicolumn{1}{c|}{1.342}&
\multicolumn{1}{c|}{1.366}&
\multicolumn{1}{c|}{1.415}&
\multicolumn{1}{c|}{1.464}&
\multicolumn{1}{c|}{*}&
\multicolumn{1}{c|}{*}&
\multicolumn{1}{c|}{*}\\
\cline{2-10}
 $^{1}P_{1}$&\multicolumn{1}{c|}{MNK}&
\multicolumn{1}{c|}{++}&
\multicolumn{1}{c|}{1.216}&
\multicolumn{1}{c|}{1.251}&
\multicolumn{1}{c|}{1.314}&
\multicolumn{1}{c|}{1.367}&
\multicolumn{1}{c|}{1.414}&
\multicolumn{1}{c|}{1.455}&
\multicolumn{1}{c|}{1.473}\\
&\multicolumn{1}{c|}{"}&\multicolumn{1}{c|}{all}&
\multicolumn{1}{c|}{1.285}&
\multicolumn{1}{c|}{1.304}&
\multicolumn{1}{c|}{1.342}&
\multicolumn{1}{c|}{1.381}&
\multicolumn{1}{c|}{1.419}&
\multicolumn{1}{c|}{*}&
\multicolumn{1}{c|}{*}\\
\cline{2-10}
& \multicolumn{1}{c|}{Sal}&\multicolumn{1}{c|}{++}&
\multicolumn{1}{c|}{1.235}&
\multicolumn{1}{c|}{1.274}&
\multicolumn{1}{c|}{1.343}&
\multicolumn{1}{c|}{1.403}&
\multicolumn{1}{c|}{1.454}&
\multicolumn{1}{c|}{1.499}&
\multicolumn{1}{c|}{1.519}\\
&\multicolumn{1}{c|}{"}&\multicolumn{1}{c|}{all}&
\multicolumn{1}{c|}{*}&
\multicolumn{1}{c|}{*}&
\multicolumn{1}{c|}{1.326}&
\multicolumn{1}{c|}{1.398}&
\multicolumn{1}{c|}{1.453}&
\multicolumn{1}{c|}{1.499}&
\multicolumn{1}{c|}{1.517}\\
\hline
\end{tabular}
\vspace*{.5cm}

$MW^{+}$ - $\omega_{0}=450~MeV$

* - absence of stable solution
\newpage
Table 1.b  The mass spectrum (in GeV) for $(u\bar{c})$ system
\vspace*{.7cm}

\begin{tabular}{|c|r|c|c|c|c|c|c|c|c|}
\hline
States&Versions&$\alpha_{1}\alpha_{2}$& x=0.0 & x=0.1 & x=0.3 & x=0.5 & x=0.7
& x=0.9 & x=1.0  \\
\hline
& \multicolumn{1}{c|}{MW}&\multicolumn{1}{c|}{++}&
\multicolumn{1}{c|}{2.012}&
\multicolumn{1}{c|}{2.031}&
\multicolumn{1}{c|}{2.065}&
\multicolumn{1}{c|}{2.097}&
\multicolumn{1}{c|}{2.127}&
\multicolumn{1}{c|}{2.155}&
\multicolumn{1}{c|}{2.168}\\
&\multicolumn{1}{c|}{"}&\multicolumn{1}{c|}{all}&
\multicolumn{1}{c|}{2.062}&
\multicolumn{1}{c|}{2.069}&
\multicolumn{1}{c|}{2.085}&
\multicolumn{1}{c|}{2.104}&
\multicolumn{1}{c|}{2.131}&
\multicolumn{1}{c|}{*}&
\multicolumn{1}{c|}{*}\\
\cline{2-10}
& \multicolumn{1}{c|}{CJ}&\multicolumn{1}{c|}{++}&
\multicolumn{1}{c|}{2.018}&
\multicolumn{1}{c|}{2.037}&
\multicolumn{1}{c|}{2.073}&
\multicolumn{1}{c|}{2.107}&
\multicolumn{1}{c|}{2.138}&
\multicolumn{1}{c|}{2.167}&
\multicolumn{1}{c|}{2.181}\\
&\multicolumn{1}{c|}{"}&\multicolumn{1}{c|}{all}&
\multicolumn{1}{c|}{2.062}&
\multicolumn{1}{c|}{2.070}&
\multicolumn{1}{c|}{2.089}&
\multicolumn{1}{c|}{2.112}&
\multicolumn{1}{c|}{2.140}&
\multicolumn{1}{c|}{*}&
\multicolumn{1}{c|}{*}\\
\cline{2-10}
 $^{1}S_{0}$&\multicolumn{1}{c|}{MNK}&
\multicolumn{1}{c|}{++}&
\multicolumn{1}{c|}{2.011}&
\multicolumn{1}{c|}{2.029}&
\multicolumn{1}{c|}{2.063}&
\multicolumn{1}{c|}{2.095}&
\multicolumn{1}{c|}{2.124}&
\multicolumn{1}{c|}{2.151}&
\multicolumn{1}{c|}{2.164}\\
&\multicolumn{1}{c|}{"}&\multicolumn{1}{c|}{all}&
\multicolumn{1}{c|}{2.058}&
\multicolumn{1}{c|}{2.065}&
\multicolumn{1}{c|}{2.082}&
\multicolumn{1}{c|}{2.102}&
\multicolumn{1}{c|}{2.127}&
\multicolumn{1}{c|}{*}&
\multicolumn{1}{c|}{*}\\
\cline{2-10}
& \multicolumn{1}{c|}{Sal}&\multicolumn{1}{c|}{++}&
\multicolumn{1}{c|}{2.012}&
\multicolumn{1}{c|}{2.031}&
\multicolumn{1}{c|}{2.065}&
\multicolumn{1}{c|}{2.097}&
\multicolumn{1}{c|}{2.127}&
\multicolumn{1}{c|}{2.155}&
\multicolumn{1}{c|}{2.168}\\
&\multicolumn{1}{c|}{"}&\multicolumn{1}{c|}{all}&
\multicolumn{1}{c|}{2.011}&
\multicolumn{1}{c|}{2.030}&
\multicolumn{1}{c|}{2.065}&
\multicolumn{1}{c|}{2.097}&
\multicolumn{1}{c|}{2.127}&
\multicolumn{1}{c|}{2.154}&
\multicolumn{1}{c|}{2.167}\\
\hline
& \multicolumn{1}{c|}{MW}&\multicolumn{1}{c|}{++}&
\multicolumn{1}{c|}{2.015}&
\multicolumn{1}{c|}{2.803}&
\multicolumn{1}{c|}{2.860}&
\multicolumn{1}{c|}{2.100}&
\multicolumn{1}{c|}{2.129}&
\multicolumn{1}{c|}{2.157}&
\multicolumn{1}{c|}{2.170}\\
&\multicolumn{1}{c|}{"}&\multicolumn{1}{c|}{all}&
\multicolumn{1}{c|}{2.063}&
\multicolumn{1}{c|}{2.070}&
\multicolumn{1}{c|}{2.087}&
\multicolumn{1}{c|}{2.107}&
\multicolumn{1}{c|}{2.134}&
\multicolumn{1}{c|}{*}&
\multicolumn{1}{c|}{*}\\
\cline{2-10}
& \multicolumn{1}{c|}{CJ}&\multicolumn{1}{c|}{++}&
\multicolumn{1}{c|}{2.020}&
\multicolumn{1}{c|}{2.039}&
\multicolumn{1}{c|}{2.075}&
\multicolumn{1}{c|}{2.109}&
\multicolumn{1}{c|}{2.140}&
\multicolumn{1}{c|}{2.170}&
\multicolumn{1}{c|}{2.183}\\
&\multicolumn{1}{c|}{"}&\multicolumn{1}{c|}{all}&
\multicolumn{1}{c|}{2.063}&
\multicolumn{1}{c|}{2.072}&
\multicolumn{1}{c|}{2.091}&
\multicolumn{1}{c|}{2.114}&
\multicolumn{1}{c|}{2.143}&
\multicolumn{1}{c|}{*}&
\multicolumn{1}{c|}{*}\\
\cline{2-10}
 $^{3}S_{1}$&\multicolumn{1}{c|}{MNK}&
\multicolumn{1}{c|}{++}&
\multicolumn{1}{c|}{2.013}&
\multicolumn{1}{c|}{2.031}&
\multicolumn{1}{c|}{2.065}&
\multicolumn{1}{c|}{2.097}&
\multicolumn{1}{c|}{2.126}&
\multicolumn{1}{c|}{2.153}&
\multicolumn{1}{c|}{2.166}\\
&\multicolumn{1}{c|}{"}&\multicolumn{1}{c|}{all}&
\multicolumn{1}{c|}{2.059}&
\multicolumn{1}{c|}{2.067}&
\multicolumn{1}{c|}{2.084}&
\multicolumn{1}{c|}{2.104}&
\multicolumn{1}{c|}{2.130}&
\multicolumn{1}{c|}{*}&
\multicolumn{1}{c|}{*}\\
\cline{2-10}
& \multicolumn{1}{c|}{Sal}&\multicolumn{1}{c|}{++}&
\multicolumn{1}{c|}{2.015}&
\multicolumn{1}{c|}{2.033}&
\multicolumn{1}{c|}{2.068}&
\multicolumn{1}{c|}{2.100}&
\multicolumn{1}{c|}{2.129}&
\multicolumn{1}{c|}{2.157}&
\multicolumn{1}{c|}{2.170}\\
&\multicolumn{1}{c|}{"}&\multicolumn{1}{c|}{all}&
\multicolumn{1}{c|}{2.012}&
\multicolumn{1}{c|}{2.031}&
\multicolumn{1}{c|}{2.067}&
\multicolumn{1}{c|}{2.099}&
\multicolumn{1}{c|}{2.129}&
\multicolumn{1}{c|}{2.157}&
\multicolumn{1}{c|}{2.170}\\
\hline
& \multicolumn{1}{c|}{MW}&\multicolumn{1}{c|}{++}&
\multicolumn{1}{c|}{2.210}&
\multicolumn{1}{c|}{2.244}&
\multicolumn{1}{c|}{2.309}&
\multicolumn{1}{c|}{2.369}&
\multicolumn{1}{c|}{2.423}&
\multicolumn{1}{c|}{2.435}&
\multicolumn{1}{c|}{2.451}\\
&\multicolumn{1}{c|}{"}&\multicolumn{1}{c|}{all}&
\multicolumn{1}{c|}{2.311}&
\multicolumn{1}{c|}{2.323}&
\multicolumn{1}{c|}{2.351}&
\multicolumn{1}{c|}{2.384}&
\multicolumn{1}{c|}{2.426}&
\multicolumn{1}{c|}{2.437}&
\multicolumn{1}{c|}{2.457}\\
\cline{2-10}
& \multicolumn{1}{c|}{CJ}&\multicolumn{1}{c|}{++}&
\multicolumn{1}{c|}{2.265}&
\multicolumn{1}{c|}{2.291}&
\multicolumn{1}{c|}{2.338}&
\multicolumn{1}{c|}{2.382}&
\multicolumn{1}{c|}{2.422}&
\multicolumn{1}{c|}{2.458}&
\multicolumn{1}{c|}{2.457}\\
&\multicolumn{1}{c|}{"}&\multicolumn{1}{c|}{all}&
\multicolumn{1}{c|}{2.328}&
\multicolumn{1}{c|}{2.340}&
\multicolumn{1}{c|}{2.366}&
\multicolumn{1}{c|}{2.394}&
\multicolumn{1}{c|}{2.425}&
\multicolumn{1}{c|}{2.459}&
\multicolumn{1}{c|}{2.479}\\
\cline{2-10}
 $^{1}P_{1}$&\multicolumn{1}{c|}{MNK}&
\multicolumn{1}{c|}{++}&
\multicolumn{1}{c|}{2.251}&
\multicolumn{1}{c|}{2.274}&
\multicolumn{1}{c|}{2.318}&
\multicolumn{1}{c|}{2.358}&
\multicolumn{1}{c|}{2.394}&
\multicolumn{1}{c|}{2.427}&
\multicolumn{1}{c|}{2.443}\\
&\multicolumn{1}{c|}{"}&\multicolumn{1}{c|}{all}&
\multicolumn{1}{c|}{2.318}&
\multicolumn{1}{c|}{2.328}&
\multicolumn{1}{c|}{2.348}&
\multicolumn{1}{c|}{2.371}&
\multicolumn{1}{c|}{2.397}&
\multicolumn{1}{c|}{2.429}&
\multicolumn{1}{c|}{2.448}\\
\cline{2-10}
& \multicolumn{1}{c|}{Sal}&\multicolumn{1}{c|}{++}&
\multicolumn{1}{c|}{2.254}&
\multicolumn{1}{c|}{2.278}&
\multicolumn{1}{c|}{2.323}&
\multicolumn{1}{c|}{2.363}&
\multicolumn{1}{c|}{2.401}&
\multicolumn{1}{c|}{2.435}&
\multicolumn{1}{c|}{2.451}\\
&\multicolumn{1}{c|}{"}&\multicolumn{1}{c|}{all}&
\multicolumn{1}{c|}{2.250}&
\multicolumn{1}{c|}{2.276}&
\multicolumn{1}{c|}{2.322}&
\multicolumn{1}{c|}{2.363}&
\multicolumn{1}{c|}{2.401}&
\multicolumn{1}{c|}{2.435}&
\multicolumn{1}{c|}{2.451}\\
\hline
\end{tabular}
\vspace*{.5cm}

* - absence of stable solution
\newpage
Table 1.c  The mass spectrum (in GeV) for $(c\bar{s})$ system
\vspace*{.7cm}

\begin{tabular}{|c|r|c|c|c|c|c|c|c|c|}
\hline
States&Versions&$\alpha_{1}\alpha_{2}$& x=0.0 & x=0.1 & x=0.3 & x=0.5 & x=0.7
& x=0.9 & x=1.0  \\
\hline
& \multicolumn{1}{c|}{MW}&\multicolumn{1}{c|}{++}&
\multicolumn{1}{c|}{2.174}&
\multicolumn{1}{c|}{2.188}&
\multicolumn{1}{c|}{2.216}&
\multicolumn{1}{c|}{2.242}&
\multicolumn{1}{c|}{2.267}&
\multicolumn{1}{c|}{2.291}&
\multicolumn{1}{c|}{2.302}\\
&\multicolumn{1}{c|}{"}&\multicolumn{1}{c|}{all}&
\multicolumn{1}{c|}{2.209}&
\multicolumn{1}{c|}{2.215}&
\multicolumn{1}{c|}{2.231}&
\multicolumn{1}{c|}{2.248}&
\multicolumn{1}{c|}{2.270}&
\multicolumn{1}{c|}{*}&
\multicolumn{1}{c|}{*}\\
\cline{2-10}
& \multicolumn{1}{c|}{CJ}&\multicolumn{1}{c|}{++}&
\multicolumn{1}{c|}{2.181}&
\multicolumn{1}{c|}{2.196}&
\multicolumn{1}{c|}{2.225}&
\multicolumn{1}{c|}{2.253}&
\multicolumn{1}{c|}{2.279}&
\multicolumn{1}{c|}{2.305}&
\multicolumn{1}{c|}{2.317}\\
&\multicolumn{1}{c|}{"}&\multicolumn{1}{c|}{all}&
\multicolumn{1}{c|}{2.210}&
\multicolumn{1}{c|}{2.219}&
\multicolumn{1}{c|}{2.236}&
\multicolumn{1}{c|}{2.257}&
\multicolumn{1}{c|}{2.281}&
\multicolumn{1}{c|}{*}&
\multicolumn{1}{c|}{*}\\
\cline{2-10}
 $^{1}S_{0}$&\multicolumn{1}{c|}{MNK}&
\multicolumn{1}{c|}{++}&
\multicolumn{1}{c|}{2.171}&
\multicolumn{1}{c|}{2.185}&
\multicolumn{1}{c|}{2.212}&
\multicolumn{1}{c|}{2.238}&
\multicolumn{1}{c|}{2.262}&
\multicolumn{1}{c|}{2.285}&
\multicolumn{1}{c|}{2.296}\\
&\multicolumn{1}{c|}{"}&\multicolumn{1}{c|}{all}&
\multicolumn{1}{c|}{2.202}&
\multicolumn{1}{c|}{2.209}&
\multicolumn{1}{c|}{2.225}&
\multicolumn{1}{c|}{2.243}&
\multicolumn{1}{c|}{2.265}&
\multicolumn{1}{c|}{*}&
\multicolumn{1}{c|}{*}\\
\cline{2-10}
& \multicolumn{1}{c|}{Sal}&\multicolumn{1}{c|}{++}&
\multicolumn{1}{c|}{2.174}&
\multicolumn{1}{c|}{2.188}&
\multicolumn{1}{c|}{2.216}&
\multicolumn{1}{c|}{2.242}&
\multicolumn{1}{c|}{2.268}&
\multicolumn{1}{c|}{2.291}&
\multicolumn{1}{c|}{2.302}\\
&\multicolumn{1}{c|}{"}&\multicolumn{1}{c|}{all}&
\multicolumn{1}{c|}{2.172}&
\multicolumn{1}{c|}{2.187}&
\multicolumn{1}{c|}{2.216}&
\multicolumn{1}{c|}{2.242}&
\multicolumn{1}{c|}{2.267}&
\multicolumn{1}{c|}{2.291}&
\multicolumn{1}{c|}{2.301}\\
\hline
& \multicolumn{1}{c|}{MW}&\multicolumn{1}{c|}{++}&
\multicolumn{1}{c|}{2.176}&
\multicolumn{1}{c|}{2.191}&
\multicolumn{1}{c|}{2.218}&
\multicolumn{1}{c|}{2.245}&
\multicolumn{1}{c|}{2.270}&
\multicolumn{1}{c|}{2.294}&
\multicolumn{1}{c|}{2.305}\\
&\multicolumn{1}{c|}{"}&\multicolumn{1}{c|}{all}&
\multicolumn{1}{c|}{2.210}&
\multicolumn{1}{c|}{2.218}&
\multicolumn{1}{c|}{2.232}&
\multicolumn{1}{c|}{2.250}&
\multicolumn{1}{c|}{2.273}&
\multicolumn{1}{c|}{*}&
\multicolumn{1}{c|}{*}\\
\cline{2-10}
& \multicolumn{1}{c|}{CJ}&\multicolumn{1}{c|}{++}&
\multicolumn{1}{c|}{2.183}&
\multicolumn{1}{c|}{2.198}&
\multicolumn{1}{c|}{2.227}&
\multicolumn{1}{c|}{2.255}&
\multicolumn{1}{c|}{2.282}&
\multicolumn{1}{c|}{2.307}&
\multicolumn{1}{c|}{2.319}\\
&\multicolumn{1}{c|}{"}&\multicolumn{1}{c|}{all}&
\multicolumn{1}{c|}{2.212}&
\multicolumn{1}{c|}{2.220}&
\multicolumn{1}{c|}{2.239}&
\multicolumn{1}{c|}{2.259}&
\multicolumn{1}{c|}{2.284}&
\multicolumn{1}{c|}{*}&
\multicolumn{1}{c|}{*}\\
\cline{2-10}
 $^{3}S_{1}$&\multicolumn{1}{c|}{MNK}&
\multicolumn{1}{c|}{++}&
\multicolumn{1}{c|}{2.173}&
\multicolumn{1}{c|}{2.187}&
\multicolumn{1}{c|}{2.216}&
\multicolumn{1}{c|}{2.240}&
\multicolumn{1}{c|}{2.265}&
\multicolumn{1}{c|}{2.288}&
\multicolumn{1}{c|}{2.299}\\
&\multicolumn{1}{c|}{"}&\multicolumn{1}{c|}{all}&
\multicolumn{1}{c|}{2.204}&
\multicolumn{1}{c|}{2.211}&
\multicolumn{1}{c|}{2.227}&
\multicolumn{1}{c|}{2.245}&
\multicolumn{1}{c|}{2.268}&
\multicolumn{1}{c|}{*}&
\multicolumn{1}{c|}{*}\\
\cline{2-10}
& \multicolumn{1}{c|}{Sal}&\multicolumn{1}{c|}{++}&
\multicolumn{1}{c|}{2.176}&
\multicolumn{1}{c|}{2.191}&
\multicolumn{1}{c|}{2.218}&
\multicolumn{1}{c|}{2.245}&
\multicolumn{1}{c|}{2.270}&
\multicolumn{1}{c|}{2.294}&
\multicolumn{1}{c|}{2.305}\\
&\multicolumn{1}{c|}{"}&\multicolumn{1}{c|}{all}&
\multicolumn{1}{c|}{2.174}&
\multicolumn{1}{c|}{2.189}&
\multicolumn{1}{c|}{2.217}&
\multicolumn{1}{c|}{2.244}&
\multicolumn{1}{c|}{2.270}&
\multicolumn{1}{c|}{2.293}&
\multicolumn{1}{c|}{2.305}\\
\hline
& \multicolumn{1}{c|}{MW}&\multicolumn{1}{c|}{++}&
\multicolumn{1}{c|}{2.438}&
\multicolumn{1}{c|}{2.459}&
\multicolumn{1}{c|}{2.497}&
\multicolumn{1}{c|}{2.533}&
\multicolumn{1}{c|}{2.567}&
\multicolumn{1}{c|}{2.599}&
\multicolumn{1}{c|}{2.614}\\
&\multicolumn{1}{c|}{"}&\multicolumn{1}{c|}{all}&
\multicolumn{1}{c|}{2.497}&
\multicolumn{1}{c|}{2.506}&
\multicolumn{1}{c|}{2.525}&
\multicolumn{1}{c|}{2.547}&
\multicolumn{1}{c|}{2.578}&
\multicolumn{1}{c|}{2.598}&
\multicolumn{1}{c|}{2.620}\\
\cline{2-10}
& \multicolumn{1}{c|}{CJ}&\multicolumn{1}{c|}{++}&
\multicolumn{1}{c|}{2.452}&
\multicolumn{1}{c|}{2.474}&
\multicolumn{1}{c|}{2.516}&
\multicolumn{1}{c|}{2.555}&
\multicolumn{1}{c|}{2.592}&
\multicolumn{1}{c|}{2.626}&
\multicolumn{1}{c|}{2.643}\\
&\multicolumn{1}{c|}{"}&\multicolumn{1}{c|}{all}&
\multicolumn{1}{c|}{2.499}&
\multicolumn{1}{c|}{2.511}&
\multicolumn{1}{c|}{2.539}&
\multicolumn{1}{c|}{2.564}&
\multicolumn{1}{c|}{2.595}&
\multicolumn{1}{c|}{2.627}&
\multicolumn{1}{c|}{2.645}\\
\cline{2-10}
 $^{1}P_{1}$&\multicolumn{1}{c|}{MNK}&
\multicolumn{1}{c|}{++}&
\multicolumn{1}{c|}{2.432}&
\multicolumn{1}{c|}{2.452}&
\multicolumn{1}{c|}{2.489}&
\multicolumn{1}{c|}{2.524}&
\multicolumn{1}{c|}{2.557}&
\multicolumn{1}{c|}{2.587}&
\multicolumn{1}{c|}{2.602}\\
&\multicolumn{1}{c|}{"}&\multicolumn{1}{c|}{all}&
\multicolumn{1}{c|}{2.481}&
\multicolumn{1}{c|}{2.491}&
\multicolumn{1}{c|}{2.511}&
\multicolumn{1}{c|}{2.534}&
\multicolumn{1}{c|}{2.561}&
\multicolumn{1}{c|}{2.587}&
\multicolumn{1}{c|}{2.605}\\
\cline{2-10}
& \multicolumn{1}{c|}{Sal}&\multicolumn{1}{c|}{++}&
\multicolumn{1}{c|}{2.438}&
\multicolumn{1}{c|}{2.459}&
\multicolumn{1}{c|}{2.497}&
\multicolumn{1}{c|}{2.533}&
\multicolumn{1}{c|}{2.567}&
\multicolumn{1}{c|}{2.599}&
\multicolumn{1}{c|}{2.614}\\
&\multicolumn{1}{c|}{"}&\multicolumn{1}{c|}{all}&
\multicolumn{1}{c|}{2.434}&
\multicolumn{1}{c|}{2.459}&
\multicolumn{1}{c|}{2.496}&
\multicolumn{1}{c|}{2.533}&
\multicolumn{1}{c|}{2.567}&
\multicolumn{1}{c|}{2.599}&
\multicolumn{1}{c|}{2.614}\\
\hline
\end{tabular}
\vspace*{.5cm}

* - absence of stable solution

\end{document}